\documentclass[12pt, preprint]{aastex}

\usepackage{graphicx}
\usepackage{amssymb}
\usepackage{epstopdf}
\usepackage{layout}
\usepackage{url}
\usepackage{subfigure}
\slugcomment{Accepted for Publication in {\it The Astrophysical Journal}}

\shorttitle{Time Dependent Modeling of 3C 454.3 Flare}
\shortauthors{Diltz et al.}

\begin{document}

\title{Leptonic and Lepto-Hadronic Modeling of the Nov. 2010 flare from 3C 454.3}

\author{C. Diltz\altaffilmark{1}, \& M. B\"ottcher\altaffilmark{2,1}}

\altaffiltext{1}{Astrophysical Institute, Department of Physics and Astronomy, \\
Ohio University, Athens, OH 45701, USA}
\altaffiltext{2}{Centre for Space Research, North-West University, Potchefstroom,
2520, South Africa}

\begin{abstract}

In this study, we use a one-zone leptonic and a lepto-hadronic model to investigate the multi-wavelength 
emission and the prominent flare of the flat spectrum radio quasar 3C 454.3 in Nov 2010. We perform a 
parameter study with both models to obtain broadband fits to the spectral energy distribution of 
3C 454.3. Starting with the baseline parameters obtained from the fits, we then investigate different 
flaring scenarios for both models to explain an extreme outburst and spectral hardening of 3C 454.3 
that occurred in Nov 2010. We find that the one zone lepto-hadronic model can successfully explain 
both the broadband multi-wavelength spectral energy distribution and light curves in the optical R, 
Swift XRT and Fermi $\gamma$-ray bandpasses for 3C 454.3 during quiescence and the peak of the 
Nov. 2010 flare. We also find that the one-zone leptonic model produces poor fits to the broadband 
spectra in the X-ray and high energy $\gamma$-ray band passes for the Nov. 2010 flare.

\end{abstract}

\keywords{galaxies: active --- galaxies: jets --- gamma-rays: galaxies
--- radiation mechanisms: non-thermal --- relativistic processes}

\section{Introduction}

\subsection{Blazars}

Blazars are a subclass of radio-loud active galactic nuclei (AGN) that possess collimated relativistic 
jets, oriented closely aligned with the line of sight \citep{Urry98,Schlickeiser96}. Blazars are 
characterized by their highly collimated jets, Doppler boosting, and multi-wavelength variability, 
in some extreme cases down to a few minutes \citep{Aharonian07,Albert07}. The broadband, multi-wavelength 
emission of blazars can be characterized by non-thermal continuum spectra with a broad low-frequency 
component in the radio-UV or even the X-ray range and a high frequency component that extends from 
X-rays to high energy $\gamma$-rays. In blazar modeling, it is generally accepted that the low 
frequency component is interpreted as synchrotron emission from a distribution of non-thermal 
electrons in the magnetic field of the jet. The origin of the second, high frequency component, 
is less clear. Two model paradigms are used in order to explain its origin. These two fundamentally 
different approaches are referred to as leptonic and hadronic models. For a detailed review, 
see \cite{Boettcher12}. \\

In the leptonic model for blazars, the broadband radiation is due mainly to leptons (electrons 
and positrons), while protons make little to no contribution since they are not accelerated to 
sufficient energies to produce comparable radiative output. The high frequency component of the 
spectral energy distribution is caused by the inverse Compton scattering of photons in the emission 
region with non-thermal electrons. Several different seed photon fields can be upscattered by the 
electrons to produce the broadband emission. These seed photon fields can include the synchrotron 
emission from the electrons themselves. This process is referred to as synchrotron self Compton (SSC) 
\citep{Jones68}. The seed photon fields can also include external photon fields, in a process known 
as external Compton scattering (EC). Different external photon fields, such as an accretion disk 
\citep{Dermer92,Dermer93}, a broad line region (BLR) \citep[BLR;][]{Sikora94,Blandford95}, or the 
dusty (infra-red emitting) torus \citep{Blazejowski00} surrounding the central accretion flow can 
serve as target photons that are then upscattered to high energy $\gamma$-rays. The leptonic model
of blazar emission has had success in modeling the broadband emission of many blazars. \\ 

In the hadronic model of blazars, synchrotron radiation from protons as well as photo-pion
production induced radiation is responsible for the emission of X-rays to high energy $\gamma$-rays 
\citep{Mastichiadis95,Mastichiadis05}. One-zone hadronic models have been used to reproduce the 
SEDs of both FSRQs and BL Lac objects \cite[e.g.,][]{MB92,Mannheim93,Muecke01,Yan14,Cerruti15,Petropoulou15}. 
Large magnetic fields are often necessary to produce strong synchrotron emission and to ensure that 
the proton Larmor radius is confined within the size of the emission region. Electron and proton 
synchrotron radiation represent the main target photon fields through which photo-hadronic 
interactions can take place, producing charged and neutral pions. The neutral pions decay to 
produce photons, while the charged pions decay to produce muons and muon neutrinos. The muons 
then subsequentally decay to produce electrons, positrons, and electron and muon neutrinos. 
Typically, the synchrotron emission from the intermediate particles is neglected and only the 
final decay products of the photo-hadronic interactions are considered \citep{Kelner08}. However, 
if the magnetic field of the emission region is large enough and the protons carry enough energy, 
then the synchrotron cooling timescale of the intermediate particles (muons and pions) can be shorter 
than their decay timescale and synchrotron emission from muons and pions can no longer be neglected 
\citep{Boettcher13}. Recently, a time dependent one-zone lepto-hadronic model, that includes synchrotron 
emission from both muons and pions, was used to reproduce the broadband SED of the FSRQ 3C 279 \citep{Diltz15}. 
The emission of muon and pion synchrotron radiation at VHE $\gamma$-rays represents a distinguishing 
characteristic between leptonic and hadronic models \citep{Aharonian00,Muecke03,Sol13}. \\ 

Both one-zone leptonic and hadronic models have been successful to reproduce the SEDs of blazars 
\citep[e.g.,][]{Collmar10,Cao14}. Therefore, additional diagnostics are needed in order to determine which model 
is best suited to reproduce all features observed in blazars. 
One of the more distinct signatures 
for the hadronic origin of $\gamma$-rays is the production of TeV -- PeV neutrinos by photo-hadronic 
interactions. Blazars as sources of high energy neutrinos have been explored in a number of different studies 
\citep{Halzen97,Muecke03,Kistler14,Dermer14b,Murase14}. \cite{Petropoulou15} modeled the broadband 
SEDs of six BL Lac objects using a one-zone lepto-hadronic model in order to determine the neutrino 
fluxes from each source. Studies have also been done to determine how the neutrino emission of FSRQs 
and BL Lac objects contributes to the cosmic neutrino background \citep{Dermer14b,Murase14,Padovani15}. 
Detecting neutrino flares during flaring events in different wavelength bands would represent a 
distinguishing signature for the hadronic origin of the $\gamma$-ray emission of blazars. 
Another unique characteristic that can separate the leptonic and lepto-hadronic origin of $\gamma$-ray 
emission would be the detection of a high degree of polarization from the X-rays and high energy 
$\gamma$-rays in blazar emission \citep{Zhang13,Zhang14}. \\

An additional diagnostic that can separate the leptonic and hadronic origin of blazar emission lies 
in the light curve behavior in different wavelength bands during flaring events. One-zone time dependent 
leptonic models have been used to investigate the flaring behavior of different blazars, such as 
1ES 1011+496 \citep{Weidinger14}, Mrk 421 \citep{Asano13,Asano15} and PKS 0208-517 \citep{Chen13}. 
The physical conditions in the jet of a blazar, such as the magnetic field, particle energy, and 
acceleration efficiency can be enhanced during a flaring event \citep{Diltz15}. Following changes 
in these physical conditions during a flare, particles (electrons/protons) are accelerated and lose 
energy on different timescales. This, in turn, produces substantially different variability patterns 
in different wavelength bands for both models. Comparing the light curves in selected bands between 
both models can help diagnose which model is best suited to explain all spectral and lightcurve
features. \\  

\subsection{Observations of 3C 454.3}

The flat spectrum radio quasar 3C 454.3 $(z = 0.859)$ is one of the brightest and most variable 
$\gamma$-ray sources in the night sky. 3C 454.3 exhibited many outbursts in the pre Fermi era. In 
2005, 3C 454.3 displayed a prominent optical outburst, reaching its historical maximum with $R = 12.0$, 
\citep{Villata06}. The event triggered follow-up X-ray observations with INTEGRAL \citep{Pian06} and 
Swift \citep{Giommi06}, $\gamma$-ray observations with AGILE \citep{Vercellone07} as well in the radio 
band, where an outburst with about a one year delay \citep{Villata07} was seen. After the launch of Fermi, 
in June 2008, 3C 454.3 has exhibited many large bursts in $\gamma$-rays and across the whole electromagnetic 
spectrum. 3C 454.3 displayed prominent outbursts in late 2009, April 2010 and late 2010 
\citep{Ackermann10,Bonnoli11,Raiteri11,Wehrle12}. On October 2010, observers noticed that 3C 454.3 
was undergoing a pronounced flaring near IR wavelengths \citep{Carasco10}. The flaring was followed 
up in optical, UV, X-ray and $\gamma$-ray bands \citep{Vercellone11}. In the following month, 
3C 454.3 obtained its highest flux in $\gamma$-rays, peaking around November 19-20  
\citep{Striani10,Sanchez10,Abdo11}. During this period, 3C 454.3 was extensively monitored in the 
radio, IR, optical and X-ray bands. Observations before the main $\gamma$-ray flare showed an increase 
in the form of a plateau, days before the main $\gamma$-ray flare occurred \citep{Vittorini14}. However, 
cross correlation analysis of the flares in the Fermi $\gamma$-ray band and Herschel/sub mm band 
during the Nov 2010 flare show no delays between them \citep{Wehrle12}. The excellent multiwavelength
coverage of this flaring event allows for a careful analysis to understand the conditions of the 
emission site and the physical mechanism responsible for the flare. \\ 

In this paper, we use the codes detailed in \cite{Diltz14,Diltz15} to provide leptonic and lepto-hadronic 
fits to the SED of the FSRQ 3C 454.3 to obtain a baseline parameter set. We then choose a subset of 
parameters from the baseline set and model them as Gaussian perturbations in time in order to simulate 
the flare that occurred on Nov 19-20, 2010. We compare the broadband SED of 3C 454.3 during the peak of the flare using both 
models. Once one model has successfully reproduced the SED during the peak of the flare, we produce 
flux light curves in the optical R, X-ray and Fermi $\gamma$-ray bands. We study the variability patterns 
and compare the results to the data to discern which model is best suited to explain the SED of 3C 454.3 and the 
light curves during and the Nov. 2010 flare \citep{Wehrle12}. We describe the leptonic and lepto-hadronic 
model setup in \S \ref{theory}; we present leptonic and lepto-hadronic model fits to the quiescent-state 
SED using the equilibrium solutions to both codes in \S \ref{SED_fitting}; we then investigate the 
light curves resulting from the Gaussian perturbations of the parameter sets in \S \ref{lightcurve}; 
we then discuss the implications of our fits and conclude in \S \ref{Discussion}. 
Throughout this paper, we convert redshift to luminosity distance using a $\Lambda$ CDM cosmology 
with $H_{0} = 70 \ km \ s^{-1}$, $\Omega_{m} = 0.3$, $\Omega_{\Lambda} = 0.7$, 
giving $d_{L} = 5.57 \ Gpc$ for 3C 454.3.

\section{\label{theory}Model Setup}

We consider a homogeneous, one zone model, where a power law distribution of ultra-relativistic particles, 
$Q(\gamma, t) = Q_{0}\gamma^{-q} \ H(\gamma; \gamma_{min}, \gamma_{max})$, is continuously injected into a 
spherical region of size $R$, moving along the jet with a bulk Lorentz factor $\Gamma$, embedded in a 
homogeneous, randomly oriented magnetic field of strength $B$. Here, $H(x; a, b)$ is the Heaviside function 
defined as $H = 1$ if $a \le x \le b$ and $H = 0$ otherwise. The size of the emission region is constrained 
through the observed variability time scale, $\Delta t_{var}^{obs}$, using the relation $R \leq c\Delta t_{var}^{obs} \delta/(1+z)$, 
where $z$ denotes the redshift to the source. The emission region moves relativistically along the jet axis 
with Doppler factor $\delta = 1/\Gamma(1 - \beta_{\Gamma} cos\theta)$. The emission is Doppler boosted into 
a viewing angle of size $\theta_{obs} \approx 1/\Gamma$, enhancing the luminosity by a factor of $\delta^{4}$ 
and reducing the  variability time scale in the comoving frame by a factor of $\delta^{-1}$. \\

Both models use coupled second order Fokker-Planck equations to track the time evolution of the particle 
distributions. We assume that the pitch angle scattering timescale is much smaller than the dynamic timescales. 
As a result, the particle distributions are isotropic in the comoving frame of the emission region. The Fokker-Planck equations of each particle species, $i = e^{\pm}, p, \pi^{\pm}, \mu^{\pm}$, is then written:

\begin{equation}
	\frac{\partial n_{i} (\gamma, t)}{\partial t} = \frac{\partial}{\partial \gamma}[\frac{\gamma^{2}}{(a+2) 
	t_{\rm acc}} \frac{\partial n_{i} (\gamma, t)}{\partial \gamma}] - \frac{\partial}{\partial \gamma} 
	(\dot{\gamma} \cdot n_{i} (\gamma, t)) + Q_{i}(\gamma, t) - 
	\frac{n_{i} (\gamma ,t)}{t_{\rm esc}} - \frac{n_{i} (\gamma ,t)}{\gamma t_{\rm decay}} - \frac{n_{i} (\gamma ,t)}{t_{\rm exp}}
\label{FP_Equ}	
\end{equation} 

\noindent where $t_{\rm acc}$ denotes the timescale due to stochastic acceleration, $a = v_{s}^{2}/v_{A}^{2}$ gives the square of the ratio between the shock and the Alfv\'en wave velocities, $t_{\rm esc}$ denotes the dynamical escape time scale for the particles which we parametrize as a multiple of the light crossing time, 
$t_{\rm esc} = \eta \, R/c$ where $\eta \ge 1$. The value of $\eta$ is kept as a free parameter. The term $\dot{\gamma}$ represents the total energy loss rate of a given particle. All charged particle species lose energy through synchrotron 
emission, with a cooling rate for a charged particle of mass $m_{i}$ given by the equation:

\begin{equation}
    \dot \gamma_{syn} = -\frac{c \sigma_{T} B^{2}}{6 \pi m_{e} c^{2}} \ (\frac{m_{e}}{m_{i}})^{3} \ \gamma^{2}
\label{syn_losses}    
\end{equation}

\noindent where $\sigma_{T}$ is the Thomson cross section. Particles can also lose energy through inverse Compton scattering of synchrotron photons and photons from external sources \citep{Dermer93}. This process is relevant only for electrons/positrons, as it is heavily suppressed for heavier particles (protons, pions and muons). Protons can also lose energy due to photo-pion production and adiabatic processes, see \citep{Huemmer10,Diltz15}. With adiabatic losses, protons will be diluted as the emission region expands. As a result, an additional term, $-n_{i}(\gamma, t)/t_{exp}$, is included in the proton Fokker-Planck equation to account for this. Here, $t_{exp}$, represents the expansion timescale of the emission region, see \citep{Pennanen14}. We neglect the adiabatic losses of electrons/positrons, pions and muons in our model. Unstable particles, such as charged pions and muons, decay on a given time scale, $t_{\rm decay}$, in their own rest frame. The term, $-n_{i}(\gamma, t)/\gamma t_{decay}$, in the Fokker-Planck equation represents the loss of the unstable particles due to decay in the laboratory frame of the emission region. \\

In both models, the particles interact with magnetohydrodynamic (MHD) Alfv\'en waves in the plasma. If the 
Doppler-shifted wave frequency is a constant multiple of the particle gyrofrequency in the particle guiding 
center frame, then a resonant interaction between the particle and the transverse component of the electric 
field of the MHD wave will occur \citep{Dermer96,Becker06,Dermer09}. The particle will experience either an 
accelerating or decelerating electric field in the transverse direction of motion over a fraction of the 
cyclotron period, resulting in an increase or decrease in energy. The accelerating or decelerating electric 
field causes the particle distributions to diffuse in energy, pushing particles to higher or lower energies 
in a diffusion pattern. This stochastic acceleration process typically causes the particle distributions to 
have a pronounced curvature in the energy spectrum \citep{1Schlickeiser84}. The strength of the particle 
diffusion depends on the spectral index of the MHD turbulence, $p$. A Kolmogorov, $p = 5/3$, or a Kraichnan, 
$p = 3/2$, spectrum are most often used to model MHD turbulence. In this study, we restrict the spectral index
of the turbulence to $p = 2$ to simulate hard sphere scattering between the MHD waves and the particle spectra. 
The stochastic acceleration timescale can be expressed as

\begin{equation}
    t_{acc} = \frac{2}{\pi} \ (\frac{p}{p-1}) \ \frac{t_{dyn}}{\beta_{A}^{2} \xi_{i}} \ \gamma^{2-p} 
\label{SA_time_scale}    
\end{equation}

\noindent where $t_{dyn}$ represents the dynamical timescale over the region in which turbulence generated
(which may be smaller than the entire emission region), $\beta_{A}$ represents the Alfv\'en velocity of the plasma 
normalized to the speed of light and $\xi_{i}$ represents the ratio of the magnetic field fluctuations relative to 
the background magnetic field, $\xi_{i} = |\Delta B/B|^{2}$. The stochastic acceleration timescale is independent 
of particle mass and will therefore be the same for all charged-particle species (protons, electrons/positrons, 
pions and muons). The diffusion term in Equation \ref{FP_Equ} describes the stochastic acceleration of particles 
in the quasi-linear approximation \citep{Dermer96}. For gyro-resonant interactions to occur in the quasi-linear 
regime, the magnetic field fluctuations must be much smaller than the background magnetic field, $\xi_{i} \ll 1$. 
If the energy density in the plasma waves starts to approach the energy density of the magnetic field, then the 
field becomes disordered and there exists no well defined gyro-frequency. In both models, we use a ratio 
between the acceleration timescale and the escape timescale as an input parameter. The ratio between the 
acceleration and escape timescales constrain the maximum size in which turbulence is injected for stochastic 
acceleration to occur in the quasi-linear regime, see Section \ref{Discussion}. \\

We numerically compute the solution of the Fokker-Planck equation between continuous injection and escape to 
reproduce the SED fit using the implicit Cranck-Nichelson (CN) scheme. Given the unconditional stability of 
the CN scheme, we can set the time step to any size. Therefore, to ensure quick convergence to the equilibrium 
solution of the Fokker-Planck equation, we set the size of the time step in our code initially to $\approx 10^{7}$~s. 
This time step size is considerably larger than the time scales for the loss terms, acceleration and escape 
terms in all particle Fokker-Planck equations. However, while choosing an arbitrarily large time step allows 
quick convergence to the equilibrium solution, it fails to probe evolutionary processes of the particle 
spectra that can occur on much smaller timescales, such as radiative, adiabatic cooling and acceleration 
processes. The Fokker-Planck equations are numerically solved simultaneously with a set of differential 
equations that track the time evolution of the photon fields until equilibrium is reached. \\

\subsection{\label{theory_general}Time Dependent Leptonic Blazar Model:}
\label{Lep_Blaz_Model}

The one-zone leptonic model used in this study is based on the work of \citep{Diltz14}. We consider the
continuous injection of relativistic electrons into an emission region of size $R$ with a randomly tangled 
magnetic field of strength $B$. The electrons emit broadband synchrotron radiation from the radio to UV / 
soft X-rays. Due to the emission of synchrotron radiation, electrons lose energy at a rate given by Equation 
\ref{syn_losses}. Synchrotron emission at lower frequencies starts to becomes more opaque due to synchrotron 
self absorption, resulting in a low energy cutoff below $\approx 10^{11} - 10^{12}~Hz$. The synchrotron photons 
created in the emission region can be upscattered through inverse Compton scattering by the same distribution 
of radiating electrons to produce synchrotron self Compton emission (SSC), peaking in the soft X-rays to soft 
to intermediate $\gamma$-rays \citep{Jones68}. \\

For FSRQs, a contribution from external Compton scattering is required to provide acceptable SED fits 
\citep{Ghisellini98, Hartman01}. External radiation fields surrounding the black hole can be upscattered 
by electrons in the emission region to produce high energy $\gamma$-rays that dominate the overall 
luminosity of the SED. For the external radiation fields in this study, we consider two sources. The 
first source is a geometrically thin, optically thick Shakura-Sunyaev accretion disk \citep{Shakura73}. 
The total bolometric luminosity of the accretion disk and the temperature of the innermost stable circular 
orbit are constrained through the mass of the supermassive black hole and the Eddington ratio of the disk. 
The second source is a a spherical isotropic blackbody radiation field surrounding the supermassive black 
hole, resembling a broad line region or an infra-red dusty torus \citep{Diltz14}. This approximation has 
had success in modeling the high energy $\gamma$-ray emission of FSRQs, low frequency and intermediate 
frequency peaked BL Lac objects \citep{Boettcher13}. Relativistic transformations are then applied to 
the two external radiation fields in the AGN frame to the comoving frame of the blob. The emission 
coefficients due to the Compton scattering of the external radiation fields are then evaluated, 
taking into account the angular distribution of the external photon fields in the comoving frame 
\citep{Dermer93, Dermer09}. Additional electron/positron pairs can be generated through $\gamma\gamma$
pair-production. The electron/positron pairs then give off their own synchrotron radiation and 
contribute to the optical/UV non thermal radiation. In our model, we evaluate the pair-production 
rate of the photon field in the emission region using the expressions provided by \cite{Aharonian83}. 
In our model, we take into consideration the full Klein-Nishina Compton cross section to evaluate the 
Compton emission coefficients and the loss rate due to Compton scattering \citep{Dermer09}. \\

\subsection{\label{theory_general}Time Dependent Hadronic Blazar Model:}
\label{Had_Blaz_Model}

The one-zone lepto-hadronic model considered in this study is based on the work of \citep{Diltz15}. 
We consider a continuous injection of relativistic electrons and protons into the emission region of 
size $R$ and magnetic field strength $B$. Large magnetic fields are necessary for protons to produce 
significant synchrotron radiation in the broadband SED. High $B$ fields are also needed to ensure that 
the proton Larmor radius is confined to within the size of the emission region, 
$R \approx 10^{15}-10^{16} \ cm$. Following the initial injection, the electrons 
and protons give off synchrotron radiation from the radio to high energy $\gamma$-rays 
and lose energy at a rate given by Equation \ref{syn_losses}. The time evolution of the 
electrons/positrons and protons are modeled through separate Fokker-Planck equations. 
The proton Fokker-Planck equation incorporates losses due to synchrotron, pion-production 
and adiabatic processes, see Equation \ref{FP_Equ}. With the proton distribution and the 
seed photon fields generated from the synchrotron emission, we evaluate the pion production 
rates based on the photohadronic interaction cross section between protons and photons. The 
total proton-photon interaction cross section is divided into separate components, corresponding 
to different channels through which the neutral and charged pions are produced: resonances (such 
as the $\Delta$ resonance), direct (non-resonant) production and multi-pion production 
\citep{Huemmer10}. We assume the target photon field for photo pion production is isotropic 
in the emission region. This limits the model to consider only photon fields that are produced 
in the emission region (synchrotron radiation). Incorporating external photon fields in the 
pion production rates would require an additional integration of the differential cross section 
against the proton and photon angular distributions. \\

We track the time evolution of the charged pions through a separate Fokker-Planck equation. 
The pions are subjected to synchrotron losses, escape and stochastic acceleration. There is 
an additional loss term in the pion Fokker-Planck equation due to pion decay. 
The pion decay timescale in the particle rest frame is $t_{decay} \approx 10^{-8} \ s$. Because of 
the short decay time for the neutral pions, $t_{decay} \approx 10^{-17} \ s$, we assume the neutral 
pions decay instantaneously. An injection term is used for a separate muon Fokker-Planck equation 
from the decay of charged pions. The muons follow their own Fokker-Planck equation, losing energy 
due to synchrotron losses and gaining energy from stochastic acceleration. The muon decay time scale 
in the laboratory frame is longer than for charged pions, $t_{decay} \approx 10^{-6} \ s$. As a result, 
charged muons can produce more synchrotron radiation before they decay compared to charged pions. The 
high energy protons must meet certain constraints with the magnetic field before pions and muons are 
energetic enough to produce synchrotron emission before decaying. If the muon and pion synchrotron 
radiation timescales are shorter than the decay time scale in the comoving frame and the photo-pion 
losses are comparable to the proton synchrotron losses, then pion and muon synchrotron emission can 
no longer be neglected, \citep{Boettcher13}. The code of \citep{Diltz15} allows us to choose arbitrary 
values of the maximum proton Lorentz factor and magnetic field, including the regime

\begin{eqnarray}
    B\gamma_{p} \geq 
        \left\{
            \begin{array}{lr}
                7.8 \times 10^{11} & \mbox{for pions}, \\
                5.6 \times 10^{10} & \mbox{for muons}.
            \end{array}
        \right. 
\end{eqnarray}     

\noindent in which muons and pions produce substantial synchrotron emission. Charged muons then decay 
to produce electron/positron pairs. A loss term in the muon Fokker-Planck equation due to charged muon 
decay serves as an additional injection term of electrons and positrons. The electrons/positrons 
generated through pair production, muon decay and from the original power law injection represent 
the source term for the electron/positron Fokker-Planck equation. The four coupled Fokker-Planck 
equations are solved simultaneously with the differential equations that model the time evolution 
of the synchrotron photon fields for each particle species: protons, charged pions, charged muons 
and electrons/positrons until the systems reach equilibrium. Bethe-Heitler pair production is not 
expected to play an important role in our model since the pion-production rate dominates over the 
loss rate from BH pair production, so it is therefore neglected. Inverse Compton scattering is also 
neglected in our model since the large magnetic field strengths in the emission region make inverse 
Compton scattering less efficient for all particle species. \\

\section{\label{SED_fitting}Comparative Modeling of 3C 454.3:}

In this section, we apply our leptonic and lepto-hadronic models discussed in the previous sections to the broadband spectral energy distribution of the flat spectrum radio quasar 3C 454.3. With the broadband multi-wavelength observations carried out on 3C 454.3  between MJD 54682 - 54770 \citep{Abdo10}, a set of input parameters for both models can be constrained. For 3C 454.3, we have the following observational parameters \citep[see][for references to the observational data]{Boettcher13}. With the redshift, $z = 0.857$, the luminosity distance to the 
source can be determined, which sets the overall luminosity scale. Superluminal motions in the jet of 
3C 454.3, $\beta_{\perp. app} \approx 15$, places a lower limit on the bulk Lorentz factor, $\Gamma \ge 15$.
The bulk Lorentz factor constrains the angle of the observer relative to the jet axis, $\theta_{obs} \le 1/\Gamma$. 
The observed variability time scale, $\Delta t_{var}^{obs} \approx 1 \ day$, constrains the size of the emission region, 
$R \le c \Delta t_{var}^{obs} \delta /(1+z)$. The mass of the supermassive black hole, 
$M_{BH} \approx 2.0 \times 10^{9} \ M_{\odot}$ \citep{Bonnoli10}, and the accretion disk luminosity, 
$L_{disk} \approx 2.6 \times 10^{46} \ erg \ s^{-1}$ \citep{Jorstad13}, allows the accretion disk 
spectral component to be determined. The measured luminosity of the broad line region (BLR), 
$L_{BLR} \approx 2.5 \times 10^{45} \ erg \ s^{-1}$, allows the size of the assumed spherical 
BLR to be estimated, $R_{BLR} \approx 0.25~pc$ \citep{Bentz13}. We approximate the BLR spectrum
as a thermal blackbody with a temperature of $T_{BLR} \approx 6.0 \times 10^{4} \ K$, which peaks 
in the UV.  \\ 

From the set of constrained input parameters, we use a "fit-by-eye" method to determine the remaining 
set of parameters for both models. The unconstrained parameters are adjusted until a reasonable fit is 
obtained. The spectral shape and normalization of the multi-wavelength emission in different wavelength 
bands allows the spectral shape and the normalization of the particle distributions to be determined. 
The magnetic field is also adjusted to provide reasonable fits to the peaks for the synchrotron, SSC 
and EC components for the one-zone leptonic model. For the lepto-hadronic model, we require that the 
proton distribution and the magnetic field are constrained in such a way so that muon and pion 
synchrotron emission can not be neglected in the SED fitting. We aim to set the input parameters in 
such a way that approximate equipartition between the particle energy and the energy density of the 
magnetic field is achieved for both models. If the relativistic jets of AGN are powered by the rotational 
energy of the central black hole, the jet will be initially Poynting-flux dominated before the magnetic 
field converts its energy into particle kinetic energy. This conversion is expected to stop at approximate 
equipartition. \\

\subsection{Leptonic Model Fits:}

Figure \ref{leptonic_flare_fit} shows the SED of 3C 454.3 and our leptonic model fit. The input parameters 
used for the leptonic fit are given in Table \ref{input_parameters}. We find that the leptonic fitting is 
satisfactory for the quiescent state for the FSRQ 3C 454.3. The IR/optical/UV emission is well explained 
by synchrotron emission from the electrons/positrons. Magnetic field values of $B \approx 1~G$ are needed 
to produce satisfactory fits for our leptonic model, consistent with previous findings for magnetic field s
trength for one-zone leptonic models \citep{Finke10,Boettcher13,Dermer14a}. The synchrotron emission becomes 
opaque at longer wavelengths due to synchrotron self absorption, producing a cutoff below $10^{12}~Hz$. 
The observed radio emission seen is due to more extended regions beyond the zone modeled that have had longer 
time to emit synchrotron radiation. We also find from the fits that the soft to intermediate X-rays are best 
explained by synchrotron self Compton emission, in accordance with leptonic modeling. While there are no 
observational features in the SED indicative of a thermal component from the disk, we include the spectral 
component of the accretion disk for completeness. \\

From the variability time scale, we can constrain the location of the emission region to be around 
$R_{\rm axis} = 2\Gamma^{2} c \Delta t_{var}^{obs}/(1+z) \approx 0.12 \ pc$. This places the emission zone within 
the broad line region, located at $R_{BLR} \approx 0.25~pc$. Using the location of the emission region, 
we compute the external Compton spectral components from the accretion disk and the BLR using the full 
Compton cross section. We find that an extensive contribution from inverse Compton scattering of the 
external radiation fields (accretion disk and BLR) is necessary to explain the low to high energy 
$\gamma$-ray emission in the quiescent state of the blazar 3C 454.3. Below a spectral break centered 
at $2~GeV$, the $\gamma$-ray emission is explained by external Compton scattering from the accretion 
disk, while above the break it is almost entirely from the Compton scattering of the BLR. 
The combination of the two external Compton components is able to reproduce the spectral 
break observed at $2 \ GeV$, consistent with previous leptonic modeling of 3C 454.3 \citep{Finke10, Boettcher13}. 
The input parameters imply a accretion disk luminosity of $L_{d} \approx 10^{46}~erg/s$ and a luminosity 
of an external radiation field of $L_{BLR} \approx 2.0 \times 10^{45}~erg/s$ to provide a satisfactory 
fit to the quiescent-state SED. These values are in accordance with observations and previous leptonic 
model fits for 3C 454.3 using a one-zone leptonic model \citep[e.g.][]{Finke10, Jorstad13}. 
Relatively long escape time scales are needed, $t_{esc} \approx 10^{7}~s$, to ensure that electrons 
have enough time to cool to produce the low energy tail of the EC disk component dominating at hard 
X-rays / soft $\gamma$-rays.  \\

\begin{table}[ht]
\centering
\begin{tabular}{cccc}
\hline
Parameter & Symbol & Leptonic Value & Hadronic Value \\
\hline
$ Magnetic \ field $ & $ B $ & $1.5$~G & $125$~G \\ 
$ Radius \ of \ emission \ region $ & $ R $ & $2.51 \times 10^{16}$~cm & $2.51 \times 10^{16}$~cm \\
$ Multiple \ for \ escape \ time \ scale $ & $\eta$ & 15 & 15 \\
$ Bulk \ Lorentz \ factor $ & $\Gamma$ & 15 & 15 \\
$ Observing \ angle $ & $\theta_{\rm obs}$ & $6.66 \times 10^{-2}$~rad & $6.66 \times 10^{-2}$~rad \\
$ Proton \ injection \ minimum \ energy $ & $ \gamma_{\rm p, min} $ & - & $ 1.0 $ \\
$ Proton \ injection \ maximum \ energy $ & $ \gamma_{\rm p, max} $ & - & $ 4.85 \times 10^{8}$ \\
$ Proton \ injection \ spectral \ index $ & $q_{p}$ & - & 2.25 \\
$ Proton \ injection \ luminosity $ & $L_{\rm p, inj}$ & - & $3.75 \times 10^{46}$~erg~s$^{-1}$ \\
$ Electron \ injection \ minimum \ energy $ & $ \gamma_{\rm e, min} $ & $ 9.0 \times 10^{2}$ & $ 5.0 \times 10^{1} $ \\
$ Electron \ injection \ maximum \ energy $ & $ \gamma_{\rm e, max} $ & $ 6.0 \times 10^{4}$ & $ 2.5 \times 10^{3} $ \\
$ Electron \ injection \ spectral \ index $ & $q_{e}$ & 2.9 & 2.9 \\
$ Electron \ injection \ luminosity \ $ & $L_{\rm e, inj}$ & $2.45 \times 10^{43}$~erg~s$^{-1}$ & $3.64 \times 10^{42}$~erg~s$^{-1}$ \\
$ Supermassive \ black hole \ mass $ & $M_{\rm BH}$ & $2.0 \times 10^{9} \ M_{\odot}$ & $2.0 \times 10^{9} \, M_{\odot}$ \\
$ Eddington \ ratio $ & $l_{\rm Edd}$ & $4.0 \times 10^{-1}$ & $4.0 \times 10^{-1}$ \\
$ Accretion \ disk \ luminosity $ & $L_{\rm disk}$ & $1.0 \times 10^{46}$~erg~s$^{-1}$ & $1.0 \times 10^{46}$~erg~s$^{-1}$ \\
$ Blob \ location \ along \ the \ jet \ axis $ & $R_{\rm axis}$ & $0.12~pc$ & $0.12~pc$ \\
$ Radius \ of \ broad \ line \ region $ & $R_{\rm BLR}$ & $0.25~pc$ & $0.25~pc$ \\
$ Luminosity \ of \ broad \ line \ region $ & $L_{\rm BLR}$ & $2.0 \times 10^{45}$~erg~s$^{-1}$ & $2.0 \times 10^{45}$~erg~s$^{-1}$ \\
$ Ratio \ of \ acc \ and \ esc \ time \ scales $ & $ t_{\rm acc}/t_{\rm esc} $ & $0.1$ & $4.0$ \\
$ Luminosity \ of \ magnetic \ field $ & $ L_{B} $ & $1.18 \times 10^{45}$~erg~s$^{-1}$ & $7.5 \times 10^{48}$~erg~s$^{-1}$ \\ 
$ Luminosity \ of \ electrons $ & $ L_{e} $ & $6.11 \times 10^{45}$~erg~s$^{-1}$ & $4.49 \times 10^{42}$~erg~s$^{-1}$ \\ 
$ Ratio \ of \ magnetic \ and \ electron \ luminosity $ & $ \epsilon_{Be} $ & $0.19$ & $1.67 \times 10^{6}$ \\ 
\hline
\end{tabular}
\caption[]{\label{input_parameters}Parameter values used for the leptonic and lepto-hadronic equilibrium fit 
to the quiescent-state SED of 3C 454.3.}
\end{table}

With the magnetic field strength and the energy spectrum for the electron distribution, we can compute the magnetization parameter, i.e., the ratio between the luminosities carried in magnetic fields and the 
electron kinetic energy, for which we find $\epsilon_{Be} = L_{B}/L_{e} \approx 10^{-1}$. 
This result is also consistent with previous findings for 3C 454.3 \citep[e.g.,][]{Finke10}.  \\ 

\subsection{Hadronic Model Fits:}

Figure \ref{hadronic_flare_fit} shows the SED fitting for our lepto-hadronic model used in this study. 
Table \ref{input_parameters} gives the corresponding set of input parameters used for the fitting. We also 
find reasonable fits for the lepto-hadronic model to the quiescent state of 3C 454.3. The IR/optical/UV 
emission is again explained by synchrotron emission of electrons/positrons. The electron injection requires 
low electron injection energies, $\gamma_{e, min} \approx 100$, and soft spectral indices, $q_{e} \approx 2.9$, 
to provide satisfactory fits. The soft X-ray to intermediate/hard $\gamma$-ray emission is explained by the 
synchrotron radiation from protons. The spectral fits required ultra relativistic energies for the protons, 
around $E_{p} \sim 10^{17} \ eV$ to reproduce the $\gamma$-ray spectra. The curvature of the proton synchrotron 
spectrum is able to reproduce the spectral break in the $\gamma$-rays observed at $\sim 2 \ GeV$. \\

A large magnetic field strength of $B \sim 100 \ G$ is needed in order to produce the necessary proton 
synchrotron radiation to provide adequate fits to the X-ray to $\gamma$-ray broadband emission. The 
protons then interact with the primary electron and proton synchrotron radiation and, through photo-hadronic 
interactions, produce pions and muons. With the maximum proton Lorentz factor and magnetic field strength 
used, the muon synchrotron cooling timescale becomes smaller than the decay time scale in the blob frame. 
As a result, muon synchrotron radiation becomes non-negligible and produces high energy $\gamma$-rays, 
centered at $\sim 20 \ GeV$. 
The combination of proton and muon synchrotron produces a shoulder in the high energy $\gamma$-ray spectrum around $20~GeV$, see Figure \ref{hadronic_flare_fit}. The charged pions produce their own synchrotron spectral component, but the emission is negligible in 
comparison to the muon synchrotron emission. The muons subsequently decay and produce electron/positron 
pairs that generate their own synchrotron spectral component. However, the synchrotron emission from the 
electron/positrons generated from charged muon decay is also negligible compared to the proton synchrotron.  \\

Using the magnetic field strength and the energy spectrum for the proton distribution, 
we find a magnetization parameter of $\epsilon_{Bp} = L_{B}/L_{p} \approx 1.12$. This result suggests 
that the emission region is slightly magnetically dominated, suggesting that in an originally Poynting-flux
dominated outflow, magnetic energy may have been efficiently converted into particle kinetic energy. This 
result contrasts with the leptonic one-zone fit for 3C 454.3 in which the emission region is more particle 
dominated.  \\

\section{\label{lightcurve}Modeling the lightcurves of 3C 454.3:}

The success of both one-zone leptonic and lepto-hadronic models in reproducing the broadband SEDs of blazars makes it difficult 
to determine the origin of the high energy $\gamma$-ray emission observed. However, blazars exhibit strong flaring across the 
electromagnetic spectrum, in which the different spectral components change and evolve on different timescales. Blazar flares 
have often been attributed to the development of strong internal shocks taking place within the jet \citep{Joshi11,Saito15}. 
The development of a strong shock can have a profound effect on the physical conditions in the jet, accelerating particles to 
extreme energies. Detailed particle-in-cell (PIC) and MHD simulations have shown that magnetic reconnection can serve as a 
viable acceleration mechanism to produce non-thermal particles in the downstream region of a strong shock \citep{Hoshino12,Sironi14}. 
Magnetic energy dissipation can also be enhanced by increased turbulence produced by strong shocks. Turbulence has been shown to 
play an important role in the dynamical evolution of magnetic reconnection \citep{Matthaeus86,Lazarian99,Yokoi13}. Shock induced 
turbulence can also play an essential role in the production of non-thermal particles through the stochastic acceleration of 
particles \citep{Veltri98,Greco02,Lazarian12}. These results show close links between the magnetic field, stochastic acceleration 
timescale, particle injection luminosity, and the particle spectral index during the development of a strong shock. How these 
spectral components change and the rate at which they change can serve as a diagnostic tool to distinguish which model can best 
explain both the quiescent and flaring states exhibited by blazars. Starting with the parameter sets of the 
quiescent-state SED fits for 3C 454.3 using the leptonic and lepto-hadronic models, we apply perturbations to a subset 
of the input parameters of both models in order to reproduce the exceptional flare observed on Nov 19-20, 2010. \\ 

We begin by allowing both models to reach equilibrium for the SED fits of 3C 454.3 as described in the previous section. We then modify the time step to $\Delta t = 2.0 \times 10^{6}$~s in the comoving frame, corresponding to $\approx 1.3 \times 10^{5}~s$ in the observer's frame. This allows us to resolve light curve patterns on time scales characteristic of synchrotron cooling effects of the relativistic protons and the acceleration time scales for the individual particle distributions. However, we are unable to model shorter-term variability timescales, 
potentially caused by the synchrotron cooling of high energy electron-positron pairs generated from the decay of charged mesons, since their cooling time scales are significantly shorter than the size of the time step selected for these simulations. Therefore, we do not explore predicted variability patterns on such short time scales, as this would increase the required computational time significantly. \\ 

We choose four input parameters to modify in both models for our light curve analysis: the particle injection luminosity, the particle spectral index, the background magnetic field and the stochastic acceleration timescale. We modify the perturbations of the selected input parameters in the form of a Gaussian function in time:

\begin{equation}
    L_{inj, i}(t) = L_{inj, 0, i} + K_{L, i} \cdot e^{-(t - t_{0})^{2}/2\sigma^{2}}
\end{equation}

\begin{equation}
    q_{i}(t) = q_{i, 0} + K_{q, i} \cdot e^{-(t - t_{0})^{2}/2\sigma^{2}}
\end{equation}

\begin{equation}
    B(t) = B_{0} + K_{B} \cdot e^{-(t - t_{0})^{2}/2\sigma^{2}}
\end{equation}

\begin{equation}
    t_{acc}(t) = \frac{t_{acc, 0}}{1 + K_{t_{acc}} \cdot e^{-(t - t_{0})^{2}/2\sigma^{2}}}
\end{equation}

\noindent where the constant $K$ denotes the strength of the perturbation. The terms 
$L_{inj, 0, i}$, $q_{i, 0}$, $B_{0}$ and $t_{acc, 0}$ denote the particle injection luminosity, spectral index, background magnetic field and stochastic acceleration timescale during quiescence, respectively. The variable $\sigma$ represents the duration of the perturbation in the comoving frame and $t_{0}$ denotes the time where the perturbation peaks in our simulation. 
In our simulations, we set $t_{0} = 5.0 \sigma$. This way, for $t = 0$, the perturbed parameter values are essentially identical to their equilibrium values, since $e^{-25} \ll 1$. The value of $\sigma$ determines the rise times of the light curves. If the cooling time scale in a given bandpass is larger than $\sigma$, then the light curve will decay on a time scale of order of the cooling time scale. If, however, the cooling timescale is smaller, then the decay time scale of the light curve will be of order $\sigma$. No extensive simultaneous SED data was collected during the periods of enhanced activity that occurred both before and after the main flare on November 19-20. Attempting to model the enhanced flux states in the optical R, Swift XRT and Fermi $\gamma$-ray bands without SED data would introduce a large number of additional, unconstrained parameters, yielding little additional insight into the physical mechanism responsible for the enhanced activity. For this reason, the flare is modeled using the quiescent state as the initial condition; the periods of enhanced activity before and after the flare are neglected. \\

\subsection{Leptonic Model Variability Analysis:}

\begin{figure}[ht]
\begin{center}
\includegraphics[height=0.45\textheight]{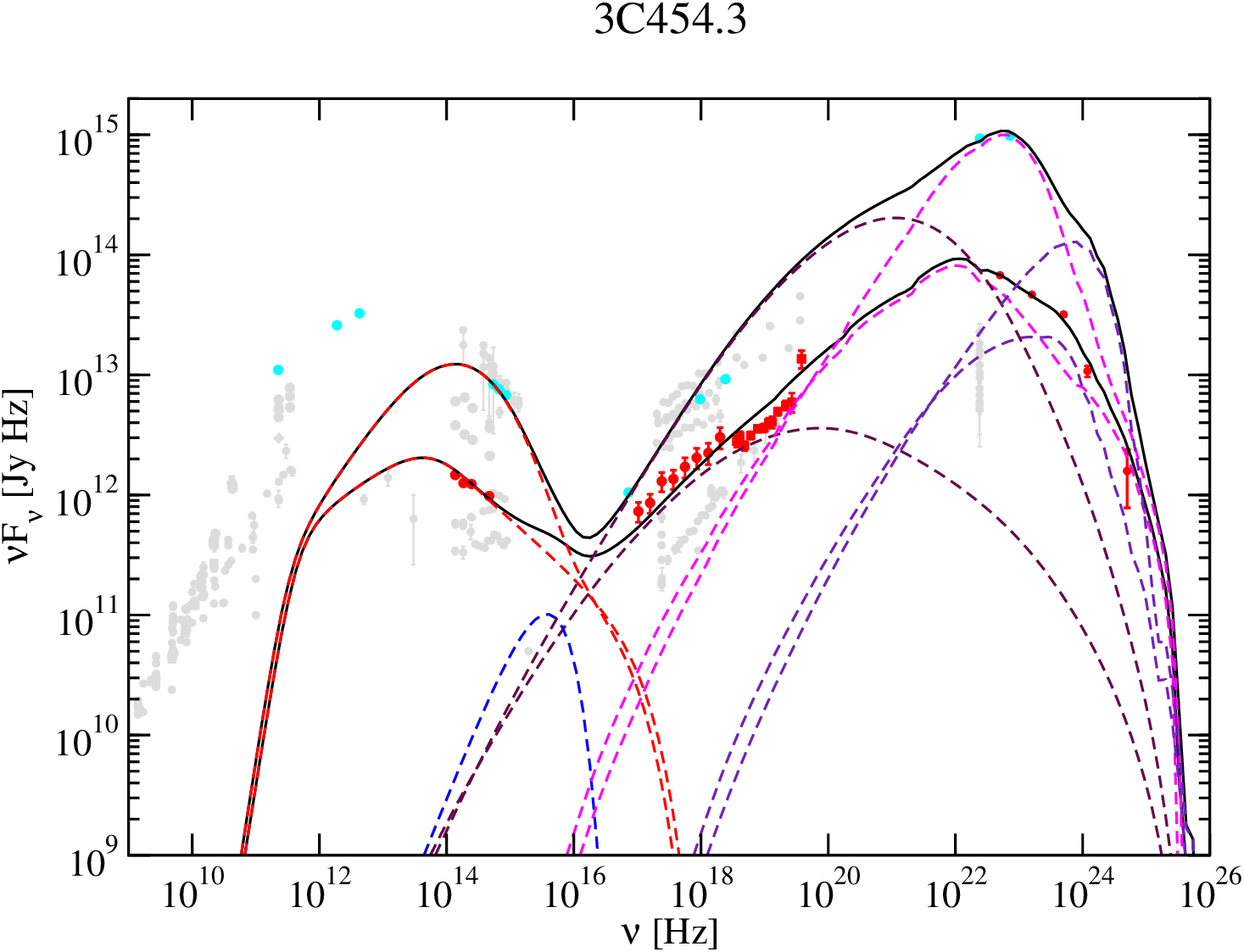}
\caption{Broadband fit to the SED of 3C 454.3 using our leptonic model during the quiescent and flaring states. The quiescent state data points included in the fit are plotted in red; additional, archival data is plotted in gray \citep[data from][]{Abdo10}. 
Broadband data during the Nov 19-20, 2010 flare are plotted in cyan \citep{Wehrle12}. The model curves are: black solid = total spectrum; red dashed = synchrotron emission from electrons/positrons; maroon dashed = synchrotron self Compton emission; blue dashed = thermal emission from accretion disk; magenta dashed
= EC emission from accretion disk; indigo dashed = EC emission from BLR.}
\label{leptonic_flare_fit}
\end{center}
\end{figure}

We now proceed to the search for a suitable combination of input-parameter variations to reproduce both the flare-state SED and the multi-wavelength light curves of 3C454.3 (see Figure \ref{leptonic_flare_fit}). The combination of perturbations has to reproduce similar levels of flux increase in the optical and HE $\gamma$-ray bands and a much weaker flux increase in the X-ray band. We begin by decreasing the acceleration time scale from its background value to $t_{acc} (t = t_{0}) = t_{acc, 0}/35$. This decrease is motivated to ensure that the acceleration timescale  drops below the cooling timescale for electrons producing synchrotron emission in the optical, $t_{cool} \approx 10^{5}~s$ in the co-moving frame. The decrease in the acceleration timescale causes the electrons to be accelerated to higher energies. As a 
result, the synchrotron emission, SSC and external Compton components are shifted to higher energies, and the SSC-dominated X-ray flux drops. Higher energy electrons also produce increased external Compton emission, see \citep{Diltz14}. The decreased acceleration timescale also changes the electron energy distribution, causing the synchrotron emission to display a pronounced curvature in the spectrum. Shifting the electrons to higher energies also causes the spectral components to become narrower since the low energy tail of the electron energy distribution is disrupted by the acceleration. The optical and high energy $\gamma$-ray 
emission increases while the X-ray emission subsequently decreases. \\

To offset the decrease in X-ray emission from altering the acceleration time scale, we increase the electron injection luminosity from $L_{inj, 0, e} = 2.5 \times 10^{43}~erg~s^{-1}$ to $L_{inj, e} (t = t_{0}) = 8.25 \times 10^{44}~erg~s^{-1}$. This subsequently causes all spectral components to increase in flux. The effect is most pronounced in the X-rays since the SSC flux scales as $n_{e}^{2}$. Increasing the electron injection luminosity also causes the non-thermal and thermal particle densities to increase. This, in turn, causes the Alfv\'en velocity of the plasma to decrease, resulting in decreased diffusion due to stochastic acceleration. The result is that both the electron synchrotron and SSC spectral components reach their peak levels for the SED, but the EC emission from the accretion disk and BLR still underpredicts the flux levels in the $\gamma$-ray bandpass. Increasing the electron injection causes the flux levels in the optical R and X-ray to overshoot  \\  

The background magnetic field is then decreased from $B_{0} = 1.5~G$ to $B(t = t_{0}) = 0.5~G$. Weakening the magnetic field inhibits synchrotron cooling for the highest energy electrons and causes them to be accelerated to even higher energies. A magnetic field decrease can represent a magnetic reconnection event where the magnetic energy is converted into particle kinetic energy. With the lower $B$ field, the synchrotron emission decreases as a result. This, in turn, causes the SSC emission to decrease as well. The higher energy electrons push the external Compton scattering of the accretion disk to higher fluxes and energies. Decreasing the magnetic field any further causes the model to over-predict the Fermi $\gamma$-ray flux, while at the same time under predicting the flux in the optical R and Swift XRT band passes. From the fits, we find that changing the spectral index of the electron injection is unnecessary. Since the electrons are already heavily accelerated, changing the electron spectral index will only increase the number of high energy electrons rather than the flux levels in the HE $\gamma$-ray band. \\

With the chosen combination of variations of the magnetic field, acceleration time scale and electron injection luminosity (see Table \ref{flare_parameters}), we obtain flux levels that are in rough agreement with the flare-state SED of 3C 454.3. However, the spectral fits in the X-ray and HE $\gamma$-rays are inconsistent with the observations. While, due to the decreased acceleration timescale, the SSC and EC components shift to higher energies, the decreased number of low energy electrons depletes the low energy SSC and EC photons, generating poor spectral 
fits to the X-ray and HE $\gamma$-ray bands during the peak of the flare. A high acceleration efficiency is necessary to upscatter photons from external sources to the flux levels of the Nov 2010 flare. Reducing the acceleration efficiency would improve the fits to the X-ray spectra, but it then underpredicts the flux levels in the HE $\gamma$-ray band during the peak of the flare. A reduced magnetic field is also necessary since weaker synchrotron cooling produces even more energetic electrons to upscatter the photons in the emission region. Increasing the magnetic field produces stronger cooling and less high energy electrons, inhibiting Compton scattering processes. The reduced synchrotron emission in the IR/optical/UV is then offset by a larger electron injection luminosity. Once the perturbations have subsided, the electrons rapidly cool by synchrotron and Compton processes. The rapid cooling produces a sharp drop in the high energy electrons responsible for the optical R and HE $\gamma$-ray emission, see Figures \ref{fermi_flare_fit} and \ref{R_band_flare_fit}. The electrons pile-up at lower energies, producing a major flare in X-rays before finally returning to the pre flux levels, see Figure \ref{XRT_flare_fit}. The strength of the perturbations used in the SED fits for our one-zone leptonic model is given in Table \ref{flare_parameters}. The resultant fits to the SED during the peak of the flare is given in Figure \ref{leptonic_flare_fit}. Based on these considerations, we conclude that the one-zone leptonic model can not reproduce the SED during the peak of the Nov 2010 flare.  \\

\subsection{Hadronic Model Variability Analysis:}

In a one-zone lepto-hadronic model, different particle distributions produce the non-thermal emission in different band passes. Changing any one of the four originally selected input parameters produces different effects to the individual particle distributions in comparison to the leptonic model. Following the same procedure as our one-zone leptonic model, we decrease the stochastic acceleration time scale to $t_{acc} (t = t_{0}) = t_{acc, 0}/4$. This acceleration time scale change applies to all particle distributions (see \ref{FP_Equ}). Since the magnetic 
field is much higher and the diffusion coefficient (responsible 
for stochastic acceleration) is larger, a modest change to the 
acceleration timescale is needed to accelerate protons to larger 
energies in comparison to the leptonic model. Decreasing the 
acceleration time scale causes low energy protons to pile up at 
higher energies, producing flares in the X-ray and HE $\gamma$-
ray bands. The higher energy protons interact with the increased proton synchrotron photon field and produce more energetic pions and muons, which then decay to produce high energy electrons/positrons. The increased amount of synchrotron radiation from both the muons and pions produces a major VHE $\gamma$-ray flare beyond $20~GeV$. Unfortunately, no observations were carried out with HESS or MAGIC that could have potentially detected flaring activity in the VHE $\gamma$-ray band. Altering the stochastic acceleration timescale produces no effect on the electron distribution responsible for the optical emission. This is due to the large magnetic field present in the emission region and the extremely short electron cooling timescales that dominate the evolution of the particle distribution in the Fokker-Planck equation. This signature is unique to lepto-hadronic models in that an X-ray and $\gamma$-ray flare can present itself while leaving the optical emission unaffected by simply changing the stochastic acceleration timescale. \\

\begin{figure}[ht]
\begin{center}
\includegraphics[height=0.45\textheight]{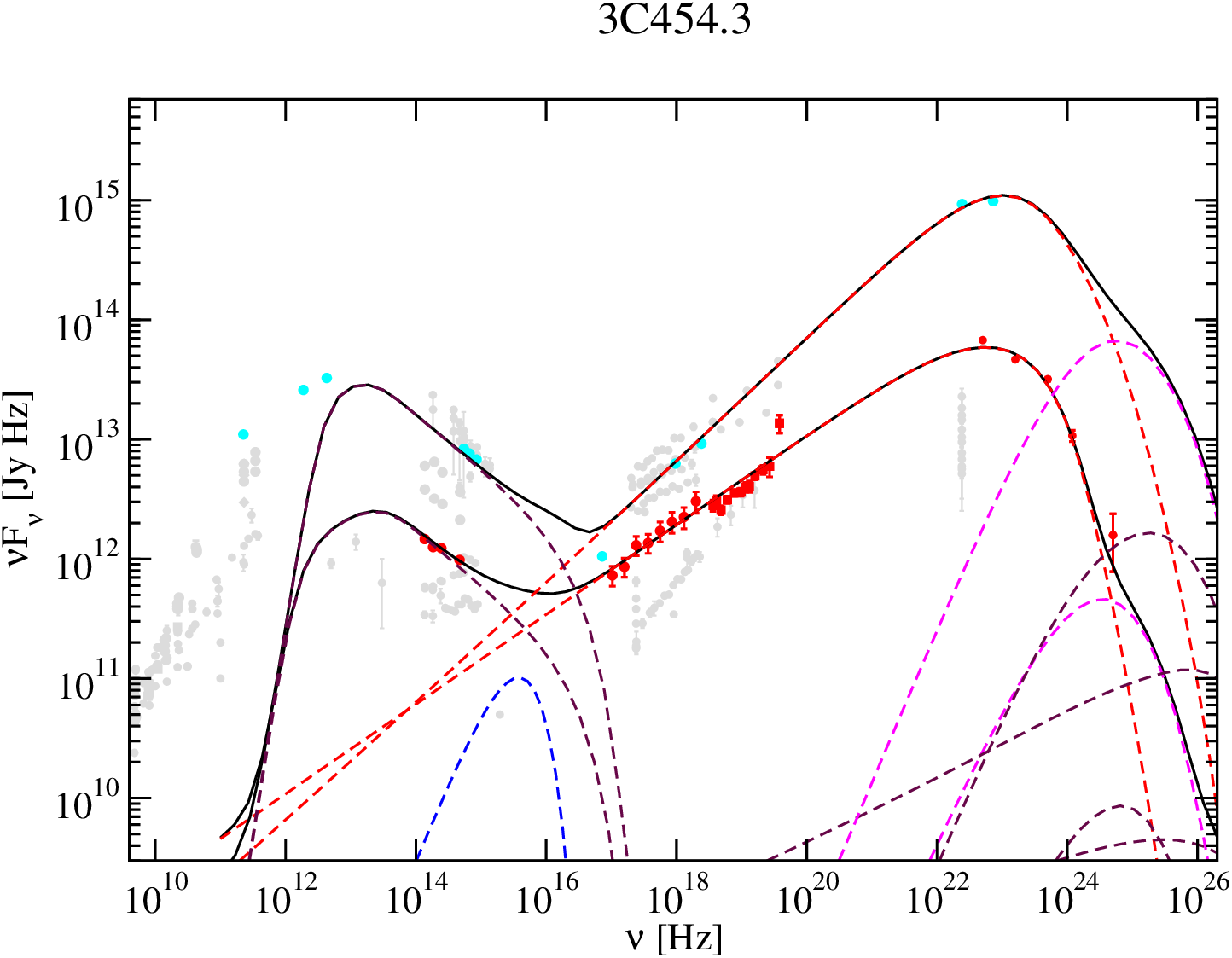}
\caption{Broadband fit to the SED of 3C 454.3 using our lepto-hadronic model during the quiescent and flaring states. The quiescent state data points included in the fit are plotted in red; additional, archival data is plotted in gray \citep[data from][]{Abdo10}. \textbf{Broadband data during the Nov 19-20, 2010 flare are plotted in cyan \citep{Wehrle12}}. The model curves are: black solid = total spectrum; red dashed = proton synchrotron emission; maroon dashed = synchrotron emission from electrons/positron; blue dashed = thermal emission from accretion disk; magenta dashed = pion synchrotron; indigo dashed = pion synchrotron.}
\label{hadronic_flare_fit}
\end{center}
\end{figure}

In order to offset the unchanged level emission from electrons, we increase the electron injection luminosity from $L_{inj, 0, e} = 3.64 \times 10^{42}~erg~s^{-1}$ to $L_{inj, e} (t = t_{0}) = 4.9 \times 10^{43}~erg~s^{-1}$. This has the effect of increasing the electron synchrotron radiation to produce a strong flare in the IR/optical/UV bands. As with the leptonic model light curve analysis of 3C 454.3, we find that it is unnecessary to change the electron spectral index during the peak of the flare. A harder or softer spectral index will worsen the spectral fits during the peak of the flare on the SED. However, changing the proton spectral index is indeed necessary to improve the fits for the harder spectral index from the X-rays to high energy $\gamma$-rays, see Figure \ref{hadronic_flare_fit}. Using a common spectral component between the X-rays and HE $\gamma$-rays and producing a harder index, from $q_{p} = 2.2$ to $q_{p} (t = t_{0}) = 1.9$ produces a strong flare in the HE $\gamma$-ray band with a moderate flare in the X-ray band. The choice of changing the proton spectral index is able to vastly improve the fits of the SED during the flare in comparison to the one-zone leptonic model. For a developing shock that is responsible for a flare, an increased compression ratio can lead to a harder spectral index, which in turn, reproduces the harder synchrotron spectra seen in the peak SED of the flare. The proton spectral index and acceleration time scale are adjusted to better improve the quality of the fits to the SED during the peak of the flare. \\ 

Due to the change in the acceleration timescale, the protons move to higher energies, but their synchrotron emission produces poor fits to the location of the observed $\gamma$-ray peak frequency during the peak of the flare. Therefore, we decreased the background magnetic field from $B = 125~G$ to $B(t = t_{0}) = 75~G$. A reduced magnetic field can be the result of magnetic reconnection events taking place in the acceleration zone. Lowering the background magnetic field leads to reduced synchrotron cooling, producing even more energetic electrons and protons. The combination of a reduced magnetic field and increased particle acceleration causes the proton synchrotron spectra to peak at $\nu_{obs}^{pk} \approx 1.5 \times 10^{23}~Hz$, consistent with the peak flux levels seen in the SED. Lowering the magnetic field has the added effect of lowering the proton and electron synchrotron flux. The acceleration time scale and electron injection luminosity are adjusted to offset the reduction in the magnetic field until a satisfactory fit is obtained for the flux levels in the broadband SED during the peak of the flare. \\ 

\begin{figure}[ht]
\begin{center}
\includegraphics[height=0.35\textheight]{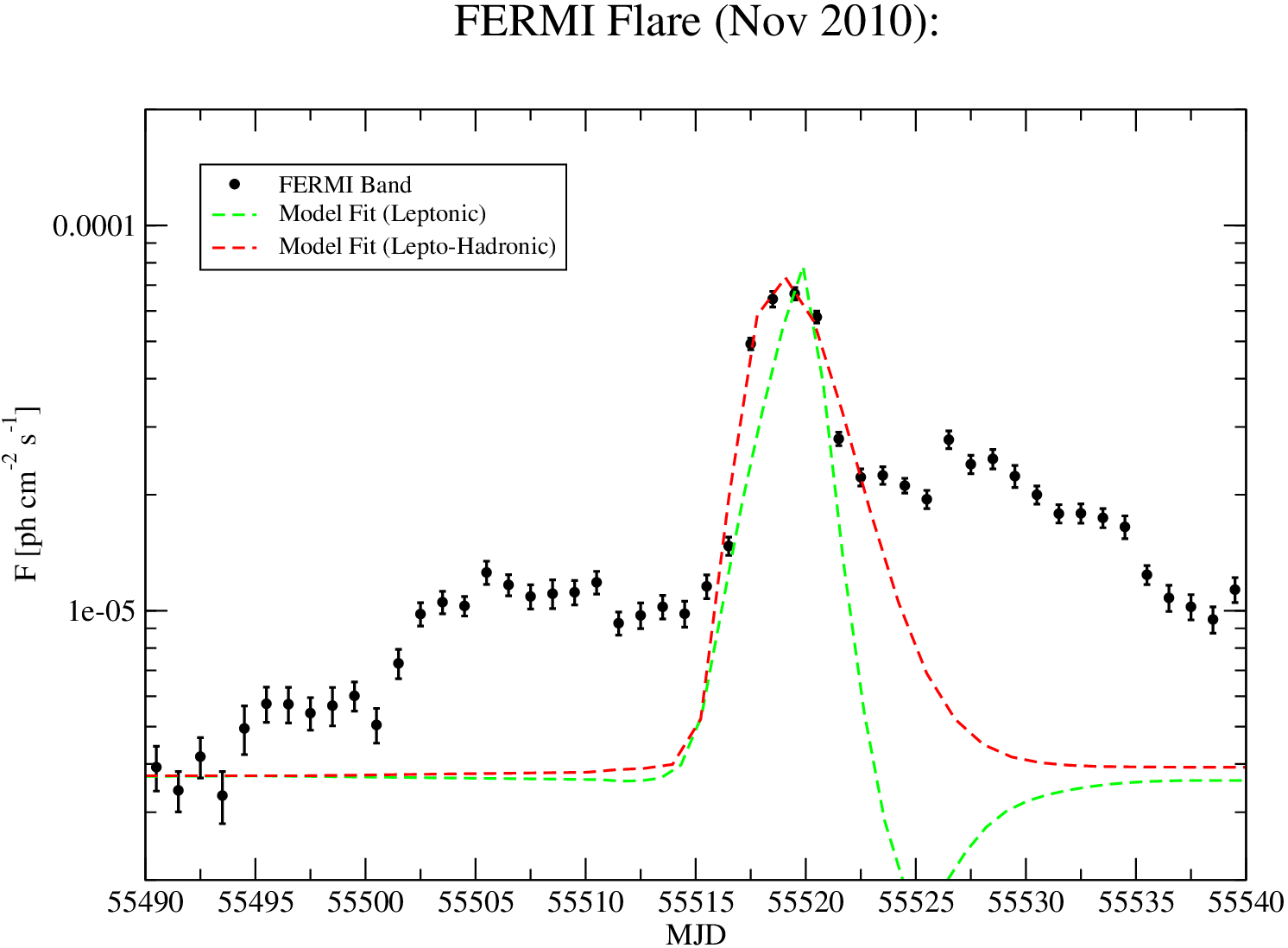}
\caption{Light curve fit between the data \citep{Wehrle12} \textbf{of the Nov 19-20, 2010 (MJD 55519-55520) flare}, and the leptonic and lepto-hadronic models in the FERMI bandpass $(20 - 300) \ GeV$ in units of $ph~cm^{-2}~s^{-1}$.}
\label{fermi_flare_fit}
\end{center}
\end{figure}

\begin{figure}[ht]
\begin{center}
\includegraphics[height=0.35\textheight]{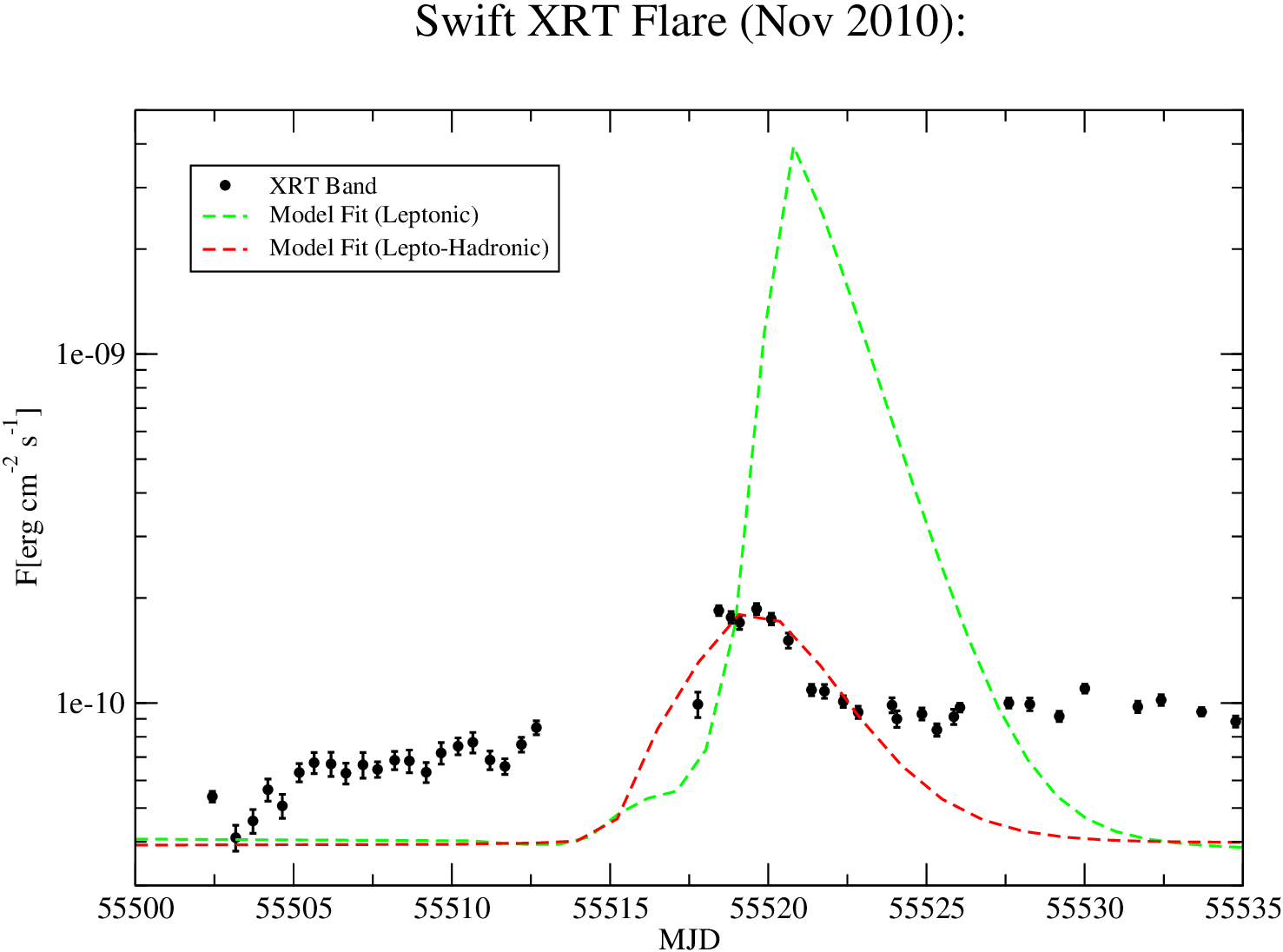}
\caption{Light curve fit between data \textbf{of the Nov 19-20, 2010 (MJD 55519-55520) flare} and the leptonic and lepto-hadronic models in the Swift XRT bandpass $(0.2 - 10) \ keV$ in units of $erg~cm^{-2}~s^{-1}$.}
\label{XRT_flare_fit}
\end{center}
\end{figure}

\begin{figure}[ht]
\begin{center}
\includegraphics[height=0.35\textheight]{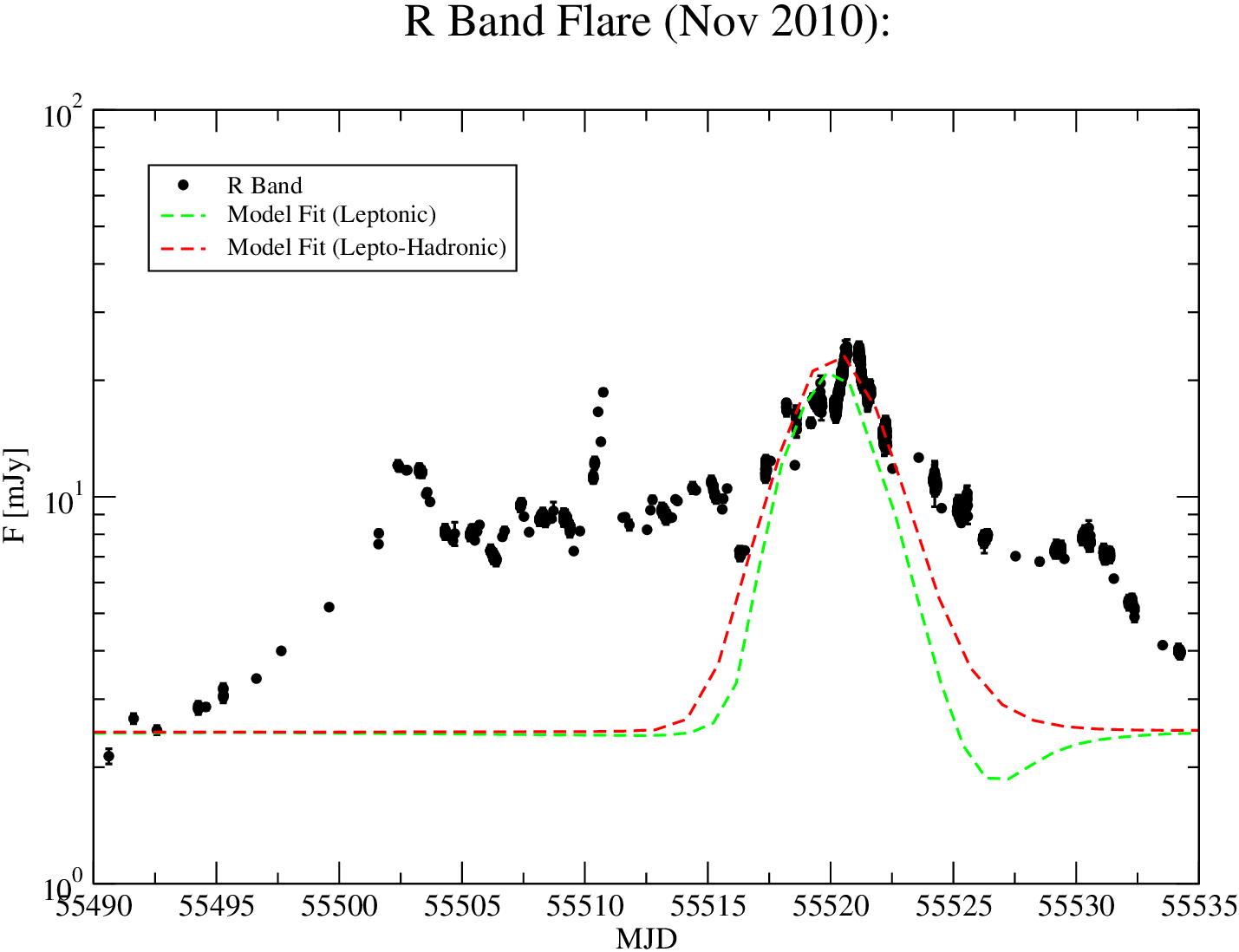}
\caption{Light curve fit between the data \textbf{of the Nov 19-20, 2010 (MJD 55519-55520) flare} and the leptonic and lepto-hadronic models in the R band in units of $mJy$.}
\label{R_band_flare_fit}
\end{center}
\end{figure}

With the variations of the magnetic field, acceleration time scale, electron injection luminosity and proton spectral index, see Table \ref{flare_parameters}, we obtain good fits to the Nov. 2010 flare SED (see Figure \ref{hadronic_flare_fit}). A shock scenario that produces a harder proton spectral index produces the spectral fits to the SED of 3C 454.3 in the X-rays and $\gamma$-rays during the peak of the Nov 2010 flare. Increasing the electron injection leads to a prominent flare in the IR/optical/UV bands that is in good agreement with the data. Decreasing both the magnetic field and acceleration time scale produces more energetic protons that reproduce the peak proton synchrotron spectra in the HE $\gamma$-rays. Based on these flare-state SED fits, we can then integrate the time-dependent fluxes over frequency in the optical R, Swift-XRT and Fermi $\gamma$-ray bands to model the light curves of the flare. The comparison shown in figures \ref{fermi_flare_fit}, \ref{XRT_flare_fit} and \ref{R_band_flare_fit} is done between model fluxes and the de-convolved measured fluxes; i.e., we have not folded our model spectra through the respective instrument response functions. The light curves in all three bands rise on a time scale of order of the shock timescale in the observer's frame $t_{rise, obs} \sim \sigma/\delta \approx 1.6 \times 10^{5}~s \approx 2~days$. As the perturbations subsides, the particles cool in the Fermi $\gamma$-ray bandpass with a synchrotron cooling timescale in the observer's frame of $t_{cool, obs} \approx 3.5 \times 10^{5}~s \approx 4~days$, consistent with the decay timescale for the light curve in the Fermi bandpass. In our model, we considered both radiative and adiabatic losses. At the highest energies, protons lose energy due to synchrotron losses, but at lower energies, adiabatic losses dominate. As the acceleration timescale drops below the adiabatic loss timescale in the comoving frame, the lower energy particles are accelerated, producing a flare in the X-ray band. Once the perturbation subsides, the particles then cool on a time scale of order of the adiabatic loss timescale, $t_{ad, obs} \approx 2.5 \times 10^{5}~s \approx 3~days$, in rough agreement with the decay timescale of the light curve in the Swift XRT bandpass during the flare. The electrons producing IR/optical/UV radiation cool via synchrotron losses at an extremely fast rate, $t_{cool, obs} \approx (1-5) \times 10^{2}~s$. As a result, the optical R band decays on a timescale similar to the shock timescale for the electron injection luminosity, $t_{decay, obs} = \sigma/\delta \approx 1.6 \times 10^{5}~s \approx 2~days$.  \\

\section{\label{Discussion}Discussion/Conclusions:}

The quiescent-state SED of 3C 454.3 can be reproduced by both a one-zone leptonic and a lepto-hadronic model. Both models suggest that the radio to IR/optical/UV emission can be explained by synchrotron emission from electrons/positrons with electron Lorentz factors of $\gamma_{e} \approx 10^{2} - 10^{3}$. Because of synchrotron self-absorption, radio emission below $10^{12}~Hz$ is strongly self-absorbed in both models. This implies that the observed radio emission must originate from a larger volume further down the jet. For the one zone leptonic model, the high energy $\gamma$-rays can be explained by a combination of external Compton scattering from the accretion disk with a luminosity of $L_{d} \approx 1.0 \times 10^{46}~erg/s$ and an external radiation field resembling a BLR with a luminosity of $L_{BLR} \approx 2.0 \times 10^{45}~erg/s$. The combination of the two spectral components is able to reproduce the spectral break observed at $2~GeV$, in accordance with previous work performed on 3C 454.3 with one-zone leptonic models \citep{Finke10,Boettcher13}. For the one-zone lepto-hadronic model, synchrotron radiation from protons with energies of $E_{p} \approx 5.0 \times 10^{17}~eV$ is responsible for the broadband emission from X-rays to high energy $\gamma$-rays. Our lepto-hadronic model considers synchrotron radiation from secondary particles, such as pions and muons, allowing us to explore larger parameter values of magnetic fields and proton energies in which pion and muon synchrotron can no longer be neglected \citep{Boettcher13,Diltz15}. The fits for the lepto-hadronic model predict the presence of a shoulder in the SED due to muon synchrotron radiation around $20~GeV$. Secondary radiation from the electron/positron pairs produced as a result of photo-hadronic interactions is negligible in comparison to muon synchrotron. Pair dominated high energy emission in blazars is typically disfavored since such models tend to over-produce the soft X-ray emission due to rapid synchrotron cooling of the electrons/positrons. \\  

In both the leptonic and the lepto-hadronic model, particles are stochastically accelerated due to gyro-resonant interactions with Alfv\'en waves with a characteristic timescale, $t_{acc}$, given by Equation \ref{SA_time_scale}. For both models, the acceleration timescale was left free as an input parameter based on the ratio between the acceleration and escape timescales in the comoving frame. The leptonic model fit suggests that the electrons are efficiently accelerated, $t_{acc}/t_{esc} = 0.1$, while the lepto-hadronic fits requires a much longer acceleration timescale, $t_{acc}/t_{esc} = 4.0$. According to Equation \ref{SA_time_scale}, the acceleration timescale is proportional to the dynamical timescale over the size of the region, $l_{dyn}$, in which turbulence is injected, $t_{dyn} = l_{dyn}/c$.  Turbulence can be injected on a broad range of size scales. The maximum size represents the size of the emission region. As the turbulence is injected, it cascades onto smaller size scales until reaching the Kolmogrov length scale where viscous dissipation of energy takes place. The size of the injection region describes the size of the eddies that propagate downstream from the onset of turbulence in MHD flows. The quasi-linear approximation assumes that the size of the magnetic field fluctuations is much smaller than the background magnetic field $\xi_{i} \ll 1$, see Equation \ref{SA_time_scale}. Typically, values near $\xi_{i} \approx 0.1 - 0.2$ represent the point where the quasi-linear approximation breaks down \citep{Zachary87}. For both model fits, the Alfv\'en velocity of the plasma falls in the range $\beta_{A} \approx 0.4 - 0.6$, and the escape timescale is $t_{esc} \approx 1.35 \times 10^{7}~s$. For the quasi-linear approximation to remain valid, the size scales of turbulent eddies that generate Alfv\'en wave turbulence, may not exceed maximum values of $l_{dyn}^{lep} \approx 6.48 \times 10^{14}~cm$ and $l_{dyn}^{had} \approx 5.82 \times 10^{16}~cm$ for the leptonic and the lepto-hadronic model, respectively.
This corresponds to $2.5 \times 10^{-2}$ and $2.15$ times the size of the emission region for the leptonic and the lepto-hadronic model, respectively. Due to the shorter acceleration timescale for the leptonic model, the turbulence needs to be generated on size scales significantly smaller than the size of the emission region for the quasi-linear approximation to remain valid. For the lepto-hadronic model, the validity of the quasi-linear approximation poses no additional constraint. \\

Large magnetic fields, $B \approx 100~G$, were necessary to provide adequate fits to the broadband SEDs with our lepto-hadronic model. These large magnetic fields are in accordance with values used in previous proton-synchrotron dominated hadronic models for FSRQs and BL Lac objects \citep{Boettcher13,Cerruti15}. 
These large magnetic field estimates are in conflict with recent measurements of radio core shifts of blazars, which argue for magnetic fields on the order of $0.1~G - 1~G$ at distances of $0.1~pc - 1~pc$ along the jet \citep{Kutkin14,1Zdziarski15}. \cite{1Zdziarski15} derived magnetic field estimates based on observed blazar jet opening angles of $\theta = k/\Gamma$, where $k = 0.1 - 0.2$ \citep{Jorstad05,Pushkarev09}. This approach led to similar magnetic field estimates based on equipartition between the energy densities in the magnetic field and in relativistic electrons, and estimates based on the flux of the partially synchrotron self-absorbed spectrum \citep{1Zdziarski15}. However, jet opening angles corresponding to $k = 1$ yield magnetic field estimates that deviate very strong from equipartition, $\epsilon_{eB} \approx 10^{-9}$. In our model, we considered jet opening angles where $\theta = 1/\Gamma$. To establish values close to equipartition, large values for the proton power, $L_{p} \approx 7.5 \times 10^{48}~erg/s$, are needed for the fits, exceeding the Eddington luminosity of the black hole by almost two orders of magnitude. These large values for the proton kinetic luminosity are a common feature in lepto-hadronic models of blazar emission \citep{Boettcher13,Cerruti15,Petropoulou15}. \\

Based on the quiescent-state SED fits, we selected a subset of the input parameters to model the prominent flare observed on Nov. 2010. We consider four parameters: magnetic field, stochastic acceleration timescale, particle injection luminosity and particle spectral index and modeled them as Gaussian functions in time (see Equations 5-8). We apply these perturbations to both models in order to determine which model can best explain both the SED of 3C 454.3 in the flare state and the light curves collected in the optical R, Swift XRT and Fermi $\gamma$-ray bands from the quiescent state SED \citep{Abdo10} to the flare state SED \citep{Wehrle12}. We found that with the one-zone leptonic model, see Table \ref{flare_parameters}, we are unable to find parameter veriations that successfully reproduce the SED in the X-ray and $\gamma$-ray bands during the peak of the flare. This is because increased stochastic acceleration (required to model the substantial $\gamma$-ray flare) depletes the number of low energy electrons responsible for the low energy tails for both the SSC and EC spectral components. Lowering the acceleration efficiency produces too few high energy electrons to upscatter the external radiation fields to the Fermi $\gamma$-ray flux levels seen during the Nov. 2010 flare. This result suggests either a temporary increase of the external radiation fields or a scenario that considers static and moving mirrors in the jet as the source of soft photons for upscattering to high energy $\gamma$-rays as a useful alternative to explain the Nov. 2010 flare \citep{Tavani15}. \\

With the one zone lepto-hadronic model, we were able to obtain acceptable spectral fits to the entire SED for both the quiescent and the flaring states.
Decreasing the spectral index of the proton distribution provides a natural explanation for the harder spectrum between X-rays and HE $\gamma$-rays seen during the peak of the flare. The model light curves resulting from our proposed flaring scenario provide satisfactory fits to the peaks of the observed optical R, Swift XRT and Fermi light curves, starting out from the quiescent state to the flare state (i.e., neglecting the plateau states before and after the flare). We did not attempt to model the enhanced states before the main Nov 19-20 $\gamma$-ray flare since no extensive simultaneous SED data was collected during this time \citep{Wehrle12}, so any such attempt would have been very poorly constrained.

\begin{table}[ht]
\centering
\begin{tabular}{lcccc}
\hline
 Scenario & $K_{L} \ [erg \ s^{-1}]$ & $K_{q}$ & $K_{B} \ [G]$ & $K_{t_{acc}}$ \\
\hline
Electron (Leptonic) & $8.0 \times 10^{44}$ & $-$ & $-0.9$ & $34.0$ \\
Proton (Lepto-Hadronic) & $-$ & $-0.3$ & $-50.0$ & $3.0$ \\
Electron (Lepto-Hadronic) & $4.5 \times 10^{43}$ & $-$ & $-50.0$ & $3.0$ \\
\hline

\end{tabular} \\ 
\caption{Model light curve fit parameters for the lepto-hadronic model. The negative value for the perturbation of the particle spectral index indicates spectral hardening. Conversely, a positive value indicates a spectral softening. $\sigma$ values are taken in the comoving frame.}
\label{flare_parameters}
\end{table}

\section*{Acknowledgments}

This work was funded by NASA through Astrophysics Data Analysis Program grant NNX12AE43G.
The work of M.B. is supported through the South African Research Chair Initiative
of the National Research Foundation\footnote{Any opinion, finding and conclusion or recommendation expressed in this material is that of the authors and the NRF does not accept any liability in this regard.} and the Department of Science and Technology
of South Africa, under SARChI Chair grant No. 64789. This study is partly based on data taken and assembled by the WEBT collaboration and stored in the WEBT archive at the Osservatorio Astronomico di Torino - INAF (http://www.oato.inaf.it/blazars/webt/).

\end{document}